\documentclass[% 
 reprint,
superscriptaddress,
nofootinbib,
 amsmath,amssymb,
 aps,
floatfix,
twocolumn,
pre,
longbibliography]{revtex4-2}

\usepackage{xcolor}
\usepackage{comment}
\usepackage{graphicx}% Include figure files
\usepackage{dcolumn}% Align table columns on decimal point
\usepackage{bm}% bold math
\usepackage{upgreek}
\usepackage{xcolor}
\usepackage{amsmath}
\DeclareMathOperator{\erf}{erf}
\usepackage[normalem]{ulem}

\definecolor{znBrown}{RGB}{200,120,0}

\begin{document}

\preprint{APS/123-QED}

\title{Basin-volume distributions in monodisperse particle packings and their role in memory formation
 %--  the soul of memory
 }

\author{Varda F. Hagh\textit{$^{\ddag}$}}
 \email{hagh@illinois.edu, \rm{Equal Contribution}}
 \affiliation{Department of Mechanical Science and Engineering, University of Illinois Urbana-Champaign, Urbana,  IL  61801,  USA}
\author{Zhaoning Liu\textit{$^{\ddag}$}}
\email{znliu@uchicago.edu,  \rm{Equal Contribution}}
\author{Sidney R. Nagel} 
    \affiliation{
 Department of Physics and The James Franck and Enrico Fermi Institutes, University of  Chicago, Chicago,  IL  60637,  USA}

\date{\today}

\begin{abstract}
Mechanically stable packings of $N$ particles in $d$ dimensions lie at the minima of an $Nd$-dimensional potential energy landscape. Starting from random initial particle positions, the system can relax using gradient-based optimization until it arrives at one of the equilibrium states; all initial conditions that end at the same minimum belong to the same catchment basin. We measure the distribution of the catchment basin volumes for indistinguishable monodisperse soft spheres in both $d=2$ and $d=3$. Ordering the basins at each system size, $N$, according to their volume, $P_N(n)$, from the largest  at $n=1$ to smaller at larger $n$, we find a very wide distribution of volumes which is similar in both dimensions: $P_N(n) \approx A_Nn^{-\alpha}$ with $\alpha \approx 1$ which, in our most favorable cases, extends over $7$ decades. We explore aspects of the connectivity of the basins, show that their structure is highly contorted, and demonstrate how these results may be used to understand the imprinting of memories in cyclic strain studies of solids.
\end{abstract}

\maketitle
\section{Introduction}

A disordered solid, such as an athermal jammed packing of soft spheres, exists in a rugged potential energy landscape with a myriad of stable configurations that defy straightforward enumeration and characterization. Since details of structure often determine function, each distinct mechanically stable arrangement corresponds to potentially different material behavior. Therefore the question naturally arises: How many stable ways are there to pack $N$ particles in a box?
This has the spirit of a ``Fermi estimation problem'', but one for which we do not have an answer; we do not know how to account for rearranging the particles to produce packings with different connectivity. 

In $d$ spatial dimensions, the $Nd$-dimensional potential energy landscape for $N$ particles consists of basins whose minima each correspond to a mechanically stable configuration. When configurations are sampled uniformly at random, the probability of finding a state is proportional to its catchment basin volume; those with larger volume will be visited more often~\cite{xu2005random,gao2009geometrical,xu2011direct,ashwin2012,frenkel2013other}. 
Such random sampling therefore not only finds stable configurations, but also measures the volume of that catchment basin -- but only if that basin can be visited multiple times.  

As $N$ or $d$ increases, the number of stable configurations grows rapidly~\cite{doye2005characterizing}. When particle permutations are considered, as is essential for polydisperse packings where particle sizes are not identical, it rapidly becomes computationally impractical to sample enough initial conditions to visit even one basin multiple times. Nevertheless, some degree of polydispersity is often regarded as important because it suppresses crystallization and promotes glass formation and the emergence of glassy behavior~\cite{kobandersen1995,o2003jamming,berthierreichman2023}. This is especially evident in two dimensions, where monodisperse packings commonly develop highly ordered domains. Consequently, many studies of basin-volume distributions have focused on systems with some particle-size dispersity~\cite{xu2005random,gao2009geometrical,xu2011direct,ashwin2012,frenkel2013other}.

In this work, we restrict our attention to packings of indistinguishable monodisperse particles where the $N!$ particle permutations sharing the same connectivity structure are treated as equivalent. This provides a great simplification by focusing only on the many fewer~\cite{doye2005characterizing,ozawa2018configurational} distinct connectivity networks that exist; it allows us to measure the distribution of the largest catchment basin volumes in packings up to $N=151$ particles in $d=3$ and up to $N=503$ in $d=2$. We show that the volume distributions are remarkably similar in both two- and three-dimensions and approximately fall into the class described by Zipf's law~\cite{zipf32} 

The use of monodisperse particles tremendously simplifies finding the distribution of basin volumes. For example, for $N=67$ particles in $3$D we found over $10^{7}$ basins that were visited multiple times, so that we know not only that these basins exist but also their absolute and relative volumes to high accuracy. Had the particles been distinguishable, as is necessarily the case for polydisperse packings, each basin would have been split into $67!$ $(>10^{94})$ sub-basins, one for each permutation of the particle labels. No single basin would then have been visited more than once in our $2\times10^{9}$ trials. Moreover, with even minute polydispersity, these sub-basins would have had slightly different energies, so energy alone could not have determined whether two packings with the same adjacency matrix are rearrangements (\textit{i.e.}, permutations) of one another or genuinely distinct packings with a different set of connections. 

Techniques that can measure the volume of a single basin, while applicable to polydisperse systems, will not in general find \textit{all} the basins that have comparable volumes. However, working with monodisperse indistinguishable particles provides a nearly complete set of basins in the region where the basins can be found multiple times. This, as we will show, is essential for being able to ask-- and answer-- some otherwise nearly intractable problems. For example, %the ability to sample nearly all of the phase space volume in many cases allows us not only to characterize the basin-volume distributions, but also to address questions that were previously inaccessible. From the essentially complete identification of over $2\times10^7$ of the largest basins, 
we can sample the volumes of basins corresponding to contiguous configurations in space; to look for correlations in their volumes. In addition, while it is difficult to measure the complex shapes of individual catchment basins, knowing the relative basin volumes over many decades allows us to identify surprising statistics of how these shapes vary among basins of similar volume.

Random sampling preferentially probes large basins, which are more likely to be sampled repeatedly, while very small basins are likely to be missed altogether. One might therefore expect this sampling bias to limit the usefulness of random sampling, since in many statistical mechanics problems the dominant contribution comes from the overwhelmingly large number of \textit{typical} configurations near the peak of the distribution rather than from the relatively rare configurations in its tail.
However, we can estimate the fraction of the total volume that we explore and find that even for relatively large $N$, our sampling covers nearly all the volume. Thus, although small basins are vastly more numerous, their combined volume is negligible; the distribution over the range we are able to measure is dominated by the configurations with large basins.

The statistics of large basin volumes also suggests an alternate interpretation of the striking formation of periodic orbits in cyclically sheared disordered solids where it only takes a modest number of cycles before the solid retains a memory of how it has been trained~\cite{Keim14, fiocco2014encoding, Royer15,lavrentovich2017period, keim2019memory, mukherji2019, arceri2021marginal, keim2021multiperiodic,lindeman2025}.  We suggest that these memories can be understood by the statistics of large basin volumes combined with the lack of correlation between basins that are contiguous in space. 

\section{Results}
\subsection{Simulations and analysis}
\textit{Simulations:} We create soft-sphere packings using periodic boundary conditions in cubic (dimension, $d=3$) or square ($d=2$) boxes of side length, $L=1$.  The particles are monodisperse with radius $R$. We use purely repulsive harmonic interactions between particles $i$ and $j$ located at positions $\mathbf{x}_i$ and $\mathbf{x}_j$. 
\begin{equation}
E_{i,j} = \ \sum_{i,j} \ \epsilon_0 \ \left( 1 - \frac{|\mathbf{x}_i - \mathbf{x}_j |}{2R} \right)^{2} \Theta \left( 1 - \frac{|\mathbf{x}_i - \mathbf{x}_j |}{2R} \right),
\label{potential}
\end{equation}
where $\epsilon_0$ sets the energy scale and $\Theta (x)$ is the Heaviside step function.  The packing fraction is $\phi = 0.85$ in $d=3$, and $\phi = 0.95$ in $d=2$. 
These are above crystalline close packing in each dimension and we find no rattlers -- particles that are unconstrained.  We choose random initial conditions for each packing (\textit{i.e.}, uniform random numbers between $0$ and $1$ for each coordinate of each particle) and sample up to $m_{\rm{tot}} \le 2\times 10^9$ starting configurations as stated for each data set. We report data for $13 \leq N  \leq 151$ in $d=3$ and $197  \leq N  \leq 509$ in $d=2$. 

The minimization procedure is described in the Methods section. There we also compare the results from two algorithms: FIRE and steepest descent. 

\textit{Determining if configurations are in the same basin:} We determine whether two initial configurations end in the same basin by first comparing the energies of minimized states. %If they are the same to $10$ decimal places, %\znliu{for the GPU based code and $10$ decimal places for the CPU based code,}  %\znliu{20 decimal places for GPU based code and 10 decimal places for CPU based code.} 
We then check to see if the packings have the same structure by comparing the connectivity of the packings with the same energy using subgraph isomorphism~\cite{juttner2018vf2++}. We did not find any cases where the energies were the same (with precision $10^{-10}$) but the connectivity was not.  Basins related by symmetries, \textit{e.g.}, reflections, rotations by $90^{\circ}$ in a square box, and rigid translations, are not counted as distinct. We then count the number of times that each distinct basin is found. 

\textit{Assigning rank-order, $n$:} For each system size, $N$, we rank the basins in descending order of the probability of being sampled, $P_N(n) \equiv m_n/m_{\rm{tot}}$ where $m_n$ is the number of times basin $n$ is found, divided by the total number of trials, $m_{\rm{tot}}$. Here, $n=1$ is the basin found the most times \textit{i.e.}, the largest basin. By construction, $P_N(n)$ monotonically decreases with increasing $n$. We note that $P_N(n)$ also represents the fraction of the total volume that the basin with rank $n$ occupies. 

\textit{Phase space statistics:} Since each of the initial $Nd$ particle coordinates could have any value between $0$ and $L=1$, the total phase space volume is 
\begin{equation}
V_{\rm{tot}} = L^{Nd}= 1.
\label{totalvolume}
\end{equation}
Therefore, the probability of landing in basin $n$ is just its volume:
\begin{equation}P_N(n)=V(n)/V_{\rm{tot}}=V(n).
\label{Pisvolume}
\end{equation}
The accuracy with which we measure a basin volume depends on the statistics of counting the number of times that the basin has been sampled: 
\begin{equation}
P_N(n) = \frac{m_n \pm (m_n)^{1/2}}{m_{\rm{tot}}}.
\label{erroronP}
\end{equation}
Below the noise floor, set by $m_n=1$, the counting error is as large as the count itself, and we no longer have any information about the shape of $P_N(n)$. We therefore define the sampled fraction of the total volume as: 
\begin{equation}
V_{\rm{samp}} = \sum_{n=1}^{n_{\rm{twice}}} V(n) = \frac{\sum_{n=1}^{n_{\rm{twice}}} m_n \pm (\sum_{n=1}^{n_{\rm{twice}}} m_n)^{1/2} }{m_{\rm{tot}}}
\label{sampledvolume}
\end{equation}
where $n_{\rm{twice}}$ is defined as the largest $n$ for which a basin was visited at least twice. 
\subsection{Three dimensions}
\begin{figure*}[t]
    \centering
    \includegraphics[width=\linewidth]{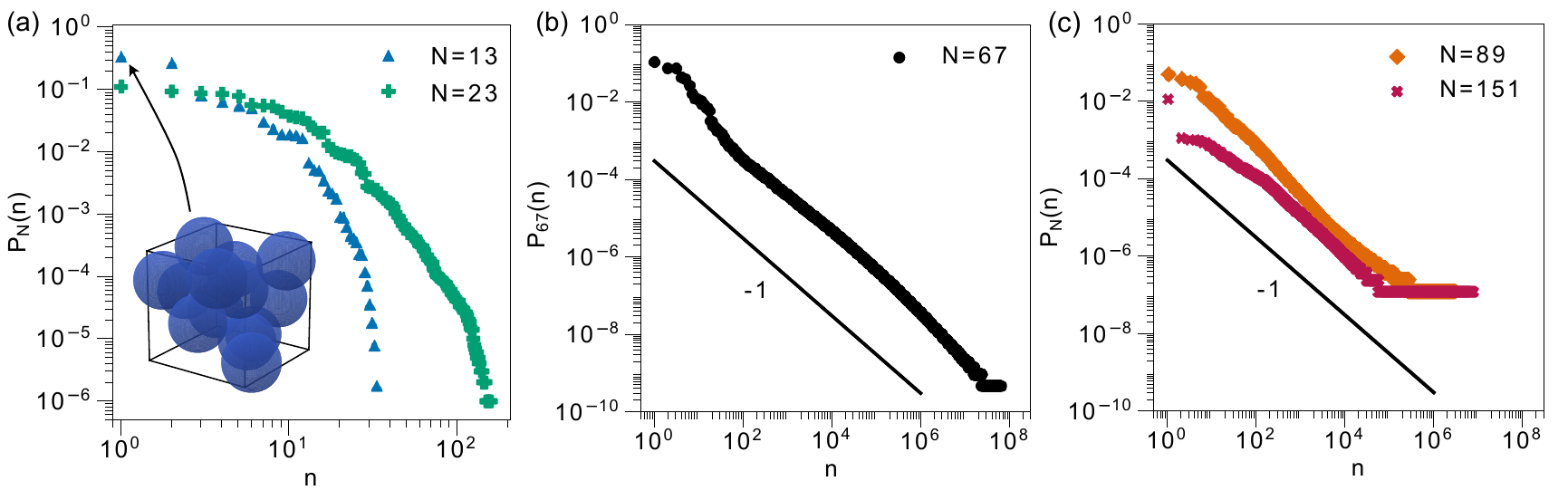}
    \caption{Distributions of catchment basin volumes in $d=3$ for systems of different sizes. (a) $P_N(n)$ for $N = 13$ and $23$. The inset shows the configuration with the largest volume ($n=1$) for $N=13$. (b) $P_{N}(n)$ for $N=67$. The total volume sampled is $V_{\rm{samp}} > 0.98$. (c) $P_N(n)$ for $N = 89$ and $151$. The total volume sampled is $V_{\rm{samp}} > 0.69$ and $V_{\rm{samp}} > 0.11$ for $N=89$ and $N = 151$ respectively. The solid lines in (b) and (c) are guides to the eye with slope $-1$. 
    }
     
    \label{fig:vol-dist}
\end{figure*}

Fig.~\ref{fig:vol-dist} shows the measured distributions $P_N(n)$ versus rank order $n$ in $d=3$ for five different system sizes $N$. For $N=13$, shown in
Fig.~\ref{fig:vol-dist}a, we find only $33$ distinct basins after $m_{\rm{tot}} = 10^{7}$ . The inset shows an image of the stable configuration with the largest basin volume (\textit{i.e.}, the most frequently found basin).  There is no guarantee that there are not other states that we did not find, but the volume of the total aggregate of those unfound states cannot greatly exceed $10^{-7}$. For $N=23$, we see similar behavior but with an expanded region before the distribution is cut off rapidly.

Fig.~\ref{fig:vol-dist}b shows data for $N=67$, sampled with $m_{\rm{tot}} > 2 \times 10^9$. In this case, many very small basins are sampled only once, while many others are presumably not sampled at all. The distribution, restricted to basins with $m_n \geq 2$, spans more than seven decades in both $n$ and volume, $V_N(n) \left(=P_N(n)\right)$. Based on the argument above, we estimate that the sampled fraction of the total volume is $V_{\rm{samp}}/V_{\rm{tot}} = V_{\rm{samp}} > 0.98$. This estimate is limited by the amount of computational time available for sampling. Although not all basins were necessarily sampled, the unsampled basins are expected to contribute negligibly to average system properties because they occupy less than $2\%$ of the total volume in phase space.

As shown in Fig.~\ref{fig:phase-space-volume}, $V_{\rm{samp}}$ increases approximately linearly with $\log m_{\rm{tot}}$ until it begins to saturate as it approaches $1$. As $N$ increases, the slope of this line decreases, making it increasingly difficult to approach $V_{\rm{samp}} = 1$. Thus, increasing $N$ affects our measurements in two ways: (i) larger systems require more time to minimize to their mechanically stable equilibrium, and (ii) many more states $n$ must be sampled in order to reach the same relative sampled volume. 

The distributions $P_{89}(n)$ and $P_{151}(n)$ in Fig.~\ref{fig:vol-dist}c have shapes similar to that of $P_{67}(n)$ in Fig.~\ref{fig:vol-dist}b, and may even be slightly straighter, with comparable power-law exponents. Moreover, the sampled volume remains reasonably large: $V_{\rm{samp}} > 0.69$ for $N=89$ and $V_{\rm{samp}} > 0.11$ for $N=151$. Even for $N=151$, this represents an appreciable fraction of the total volume: at least one in ten simulations that average over multiple starting configurations would be expected to fall into this large-basin regime. This is an important regime that cannot be disregarded as being `merely a tail' of the distribution.

We note that for $n \gtrsim 10$, $P_{67}(n)$ is relatively smooth. Although not perfect, it is reasonably approximated over the range $10 < n < 2 \times 10^7$ by a straight line on the log-log plot. Below $n \approx 10$ there are fluctuations.  These features are broadly consistent with Zipf behavior~\cite{zipf32}, 
\begin{equation}
P_{\rm{Zipf}}(n) \sim n^{-\alpha}; \ \ \alpha \approx 1.
\label{powerlaw}
\end{equation} 
However, there %are deviations from a pure power law at small $n$ and there 
is a small change in slope above $n \approx 10^6$, where the distribution begins to saturate. Fig.~\ref{fig:comparison-3D} of the Supplementary Information shows the comparison between fits derived from a power-law and log-normal volume distributions. It is not clear that either form provides a decisively better description.

For $N=67$, there may be an enormous number of exceedingly small basins that are not sampled. However, these basins occupy only a very small fraction of the total volume and therefore contribute negligibly to the volume weighted averages. Thus for determining which states contribute to average system properties in a meaningful way, the relevant feature is not whether the full distribution $P_N(n)$ is log-normal, as has been argued elsewhere~\cite{xu2011direct,asenjo2014numerical,paillusson2015devising,martiniani2017numerical,casiulis2023when,frenkel2013other}, but rather the behavior of the large-volume tail; the dominant contribution comes not from basins near the peak of the distribution, but from the tail extending to large volumes.

One might expect, consistent with trends reported for hard-WCA particles~\cite{casiulis2023when}, that the basins with the largest volumes are those whose minima have the lowest potential energy. We have examined the correlation between basin volume and energy for the $N=67$ system in $d=3$. For each basin $n$, we calculate  the total potential energy at the basin minimum: $E(n) = \frac{1}{2} \sum_{i \neq j} E_{i,j}$, where $E_{i,j}$ is obtained from Eq.~\ref{potential}. 

We plot $E(n)$ versus basin volume, $P_{\rm N}(n)=V_{\rm N}(n)$ for $N=67$, in Fig.~\ref{fig:energy-p-67} of the Supplementary Information. While there is a slight trend for lower-energy configurations to have larger catchment volumes, the correlation is, to a large extent, obscured by the large spread in $E_{\rm N}(n)$ among basins of similar volume. In Fig.~\ref{fig:energy-p-67}, the lowest value of $E_{\rm N}(n)$ is roughly constant with $P_{\rm N}(n)$, while the highest values of $E_{\rm N}(n)$ decrease gradually with increasing $P_{\rm N}(n)$. Notably, we found one basin with a tiny volume, sampled only once in the two billion trials, that has the third-lowest energy of all basins found; at the other extreme, the basin with the largest volume does not have the lowest energy, although it is among the lowest. So although our data are broadly consistent with the trends reported for hard-WCA particles, where lower-energy minima were found to have larger basin volumes~\cite{casiulis2023when}, individual basins can deviate strongly from it. The correlation holds at best on average with large fluctuations around it. 

\subsection{Two dimensions} 

The data presented for the $d=3$ catchment basin distributions suggest a very special behavior. Some basins that are found frequently are over seven orders of magnitude larger in volume than others. One might suppose that this is due to geometry, that is, to a special way in which the particles in these large basins are packed in three dimensions. Another possibility is that this behavior is due to a more general feature of high-dimensional landscapes.

\begin{figure}[t]
    \centering
    \includegraphics[width=1\linewidth]{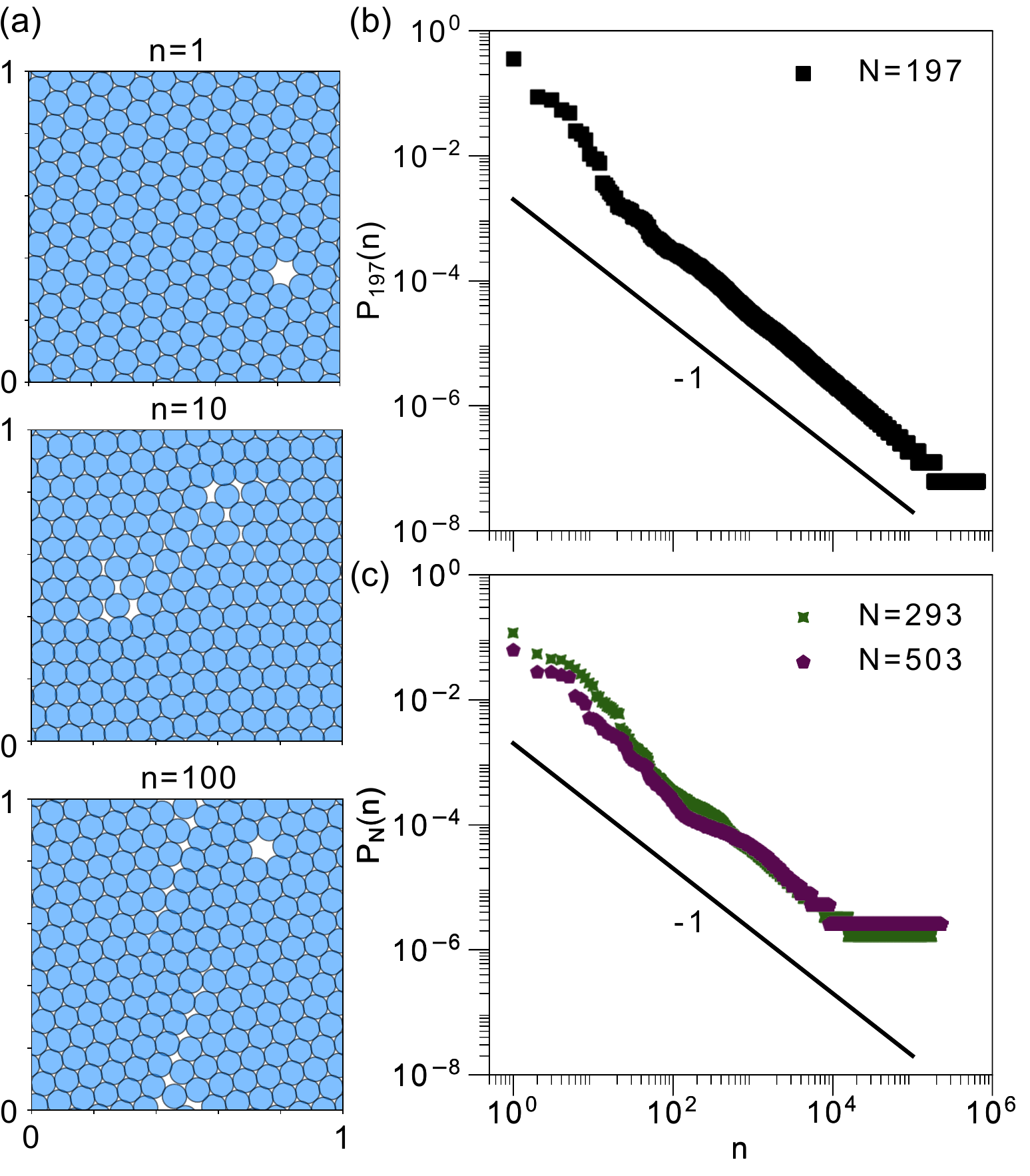}
    \caption{Images of three packings and distribution of catchment basin volumes in $d = 2$. (a) Images of three configurations with $n=1, 10, 100$ for $N=197$ showing large seemingly ordered domains. (b) $P_{N}(n)$ for $N = 197$.  The total volume sampled is $V_{samp} > 0.97$. (c) $P_{N}(n)$ for $N=293$ and $N=503$. The solid lines are guides to the eye with slope $-1$.}
    \label{fig:vol-dist-2D}
\end{figure}

We can examine these possibilities by performing an equivalent analysis of two-dimensional packings of monodisperse disks. In the $d=2$ case, we expect the structure to appear highly ordered, with large domains of nearly crystalline structure, as shown in Fig.~\ref{fig:vol-dist-2D}a. Thus, the packing geometry in $d=2$ is very different from that in $d=3$ (see inset of Fig.~\ref{fig:vol-dist}a). One might therefore expect that, if geometry is the crucial feature responsible for the $d=3$ distributions in Figs.~\ref{fig:vol-dist}b,c, then the corresponding distributions in $d=2$ would be very different.

In Figs.~\ref{fig:vol-dist-2D}b,c, we show the distribution functions for two-dimensional systems of monodisperse disks over a range of system sizes. We find that the $P_N(n)$ distributions are largely similar to those found in $d=3$.
One difference is that there are more fluctuations in $d=2$ than in $d=3$. For some values of $N$, these fluctuations produce structure in $P_N(n)$, \textit{i.e.}, deviations from a straight line in the log-log plot. However, this structure changes when $N$ is varied by very small increments, as shown in Fig.~\ref{fig:fluctuations-2D} of the Supplementary Information. There does not appear to be a systematic trend. We also show the comparison between power-law and log-normal fits to the two-dimensional distributions in Fig.~\ref{fig:comparison-2D} of the Supplementary Information.

\subsection{Consequences of the form for $P_N(n)$}

We have measured catchment basins over a very large range of volumes, in some cases (see Fig.~\ref{fig:vol-dist}b) up to a ratio $V(n=1)/V(n=n_{\rm{twice}}) \approx 10^8$, and for reasonably large system sizes. Since we have also sampled a substantial fraction of the total basin volume, $V_{\rm{samp}}$, we can address physical questions about the structure of the energy landscape that would otherwise be difficult to investigate. For example, as shown below, we can determine how volumes are correlated between contiguous basins with a shared border. As we then argue, this provides a new perspective on how memories are formed in cyclically sheared solids. These extensions are only possible because of the use of monodisperse, indistinguishable particles that allowed a nearly complete set of basins to be visited multiple times in our random sampling.

In the regime where $P_N(n) \approx A_N n^{-\alpha}$ with $\alpha \approx 1$, 
if we change variables so that we measure in terms of $y \equiv \log n$, then $P_N(y) dy = P_N(n) dn$ or $P_N(y) = P_N(n)|dn/dy| \approx A_N$. That is, the probability of finding a catchment basin of size $y=\log n$ is nearly constant, meaning that it is approximately equally likely to find a basin in any decade of $n$. Deviations from a pure power law with $\alpha=1$ will produce deviations from this expectation. We will use this result below to measure correlations between catchment basin volumes and to relate the distributions to memory formation in cyclically sheared packings. It is interesting that the exponents that we measure are close to $\alpha=1$, which marks a dividing line: if $\alpha>1$ then the distribution would be biased toward small $n$ (large volumes), while if $\alpha<1$ then  smaller basin volumes would dominate the system.

\subsection{Exploring neighboring basins}  

In order to study correlations between volumes of neighboring basins, we start with a basin, $n$, chosen from our distribution and sample some of its neighbors to determine their sizes. We cover a wide range of $n$ by choosing, as starting points, each of the first ten basins in each decade of the distribution, (\textit{i.e.}, $1 \leq n \leq 10$, $11 \leq n \leq 20$, $101 \leq n \leq 110$, $\dots$). For each chosen basin, we start at its minimum and choose a random displacement direction in the $Nd$-dimensional phase space. We then incrementally increase the displacement along that direction; after each increment, we test whether the system relaxes back to its original configuration once the forcing is removed. We repeat this process until the system falls into a new basin neighboring the original one. This procedure guarantees that the newly sampled basin is one of the nearest neighbors of the original basin and shares a border with it along the chosen displacement direction.

Once the system relaxes to the minimum of a nearest neighboring basin, we label that nearest neighbor basin as $n_{nn}$ and determine whether it is one of the basins found in our initial sampling. If it is, we record the decade of $n$ into which it falls. If the neighboring basin is not recognized as having appeared in our previous sampling, we know that its volume is smaller than our lower limit, and we assign it to the bin, ``unknown'', with $n > n_{\rm{twice}}$. For each initial basin, we repeat this process $500$ times for $N=67$ and $50$ times for $N=89$. In this way, we can determine whether the volumes of neighboring basins are correlated with the volume of the initial basin indexed by $n$.

\begin{figure}
    \centering
    \includegraphics[width=1\linewidth]{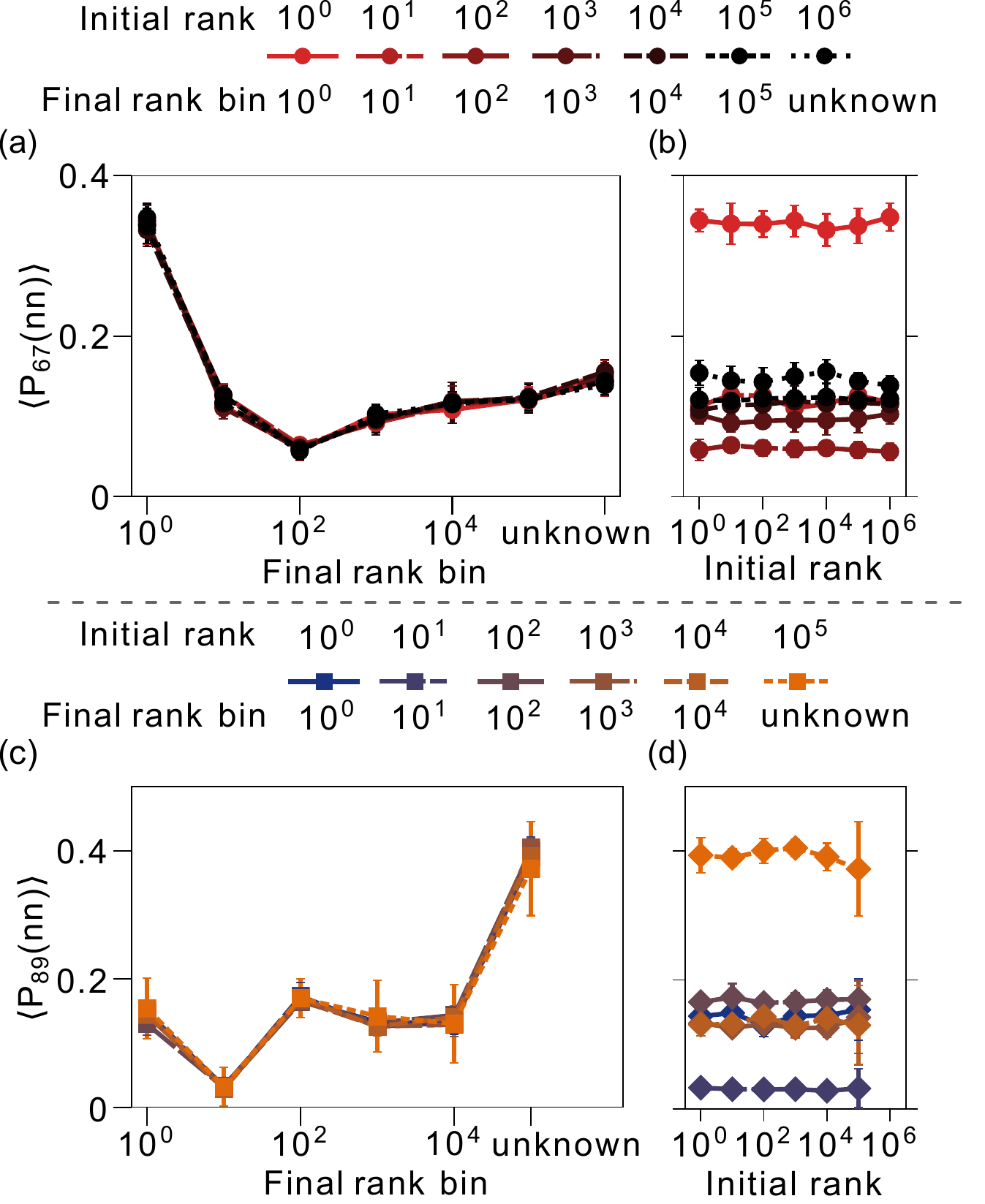}
    \caption{Distribution of nearest-neighbor basin volumes, $P_N(n_{nn})$, for systems of two sizes. (a, b) $N=67$ and (c, d) $N=89$, both in $d=3$. (a, c) Final volumes of the basins into which the system relaxes after being randomly perturbed out of its initial basin, indexed by $n$, as described in the text. Here, $P_N(n_{nn})$ is plotted versus the decade into which the neighboring basin falls. The distribution is approximately uniform across decades for both $N=67$ and $N=89$. The last bin on the right contains all basins that were not originally found more than once in the data shown in Fig.~\ref{fig:vol-dist}b, c. Its height relative to the other bins, each of which corresponds to one decade in $n$, indicates how many decades were not sampled by our initial protocol. (b, d) $P_N(n_{nn})$ plotted versus the decade from which the starting basin was chosen. The local distribution of neighboring basin volumes is uncorrelated with the initial basin volume.} 
    
    \label{fig:neighboring-basins}
\end{figure}

The results of this measurement are shown in Fig.~\ref{fig:neighboring-basins} for $d=3$ with $N=67$ and $N=89$. We find that $\langle P_N(nn)\rangle$ is approximately constant when binned logarithmically, consistent with the argument above. Most importantly, in all cases, the results are consistent with the neighboring basin volumes, $V_{nn}$, being independent of the the volume of the initial basin, $V_{n}$. The increase seen in the first decades of Figs.~\ref{fig:neighboring-basins}a and c is consistent with the first decade of the distributions $P_{67}(n)$ and $P_{89}(n)$ lying somewhat above the power law fit to the rest of those distributions. That is, if neighboring basins are sampled independently of the initial basin volume, their distribution should be nearly constant on a logarithmic scale.

The very last bin shows the sum over all the decades that extend \textit{past} the region sampled with $m_n \ge 2$ in the distributions.  We do not know the volumes of those basins, but we know that they are smaller than any basins sampled reliably. From the height of this bin, one can estimate how many unsampled decades lie beyond the region accessible to our sampling. For $N=67$, there are approximately $1.5$ unsampled decades; for $N=89$, there are approximately $3.5$ unsampled decades. These values are also consistent with our extrapolations in Fig.~\ref{fig:phase-space-volume}. There, we extrapolate the logarithmic growth of $V_{\rm{samp}}$ with $m_{\rm{tot}}$ and estimate how many additional decades would be required to reach $V_{\rm{samp}} = V_{\rm{tot}} = 1$. In such an extrapolation, however, it is difficult to account for the curvature that appears as $V_{\rm{samp}}$ saturates near $1$.

\subsection{Statistics of basin shapes: evidence of volume contortions}

We have determined $P_N(n)$ over a broad range of basin volumes, which enabled us to measure correlations between the volumes of neighboring basins. We now extend this analysis by defining and measuring a statistical quantity that captures one aspect of the geometry of individual catchment basins. Specifically, we compare the basin volumes measured via sampling with the estimated volumes of the hyper-rectangles defined by the normal modes of each basin. This comparison suggests that basin shapes are highly variable and have contorted surfaces.

If we assume that the basins are convex with no narrow or extended protrusions, we can make a crude estimate of their volumes using the $Nd$ normal mode directions of their corresponding configurations. These normal modes are the eigenvectors of the Hessian matrix, $H_{ij} = \frac{\partial^{2} E}{\partial x_{i} \partial x_{j} }$ where $E$ is the contact potential energy function. Following the work of Xu \textit{et al.,}~\cite{xuvitelli20}, for each basin, $n$, we move out along each normal mode, labeled by $\mu$ in both the positive and negative directions until the system falls into a neighboring basin as we did above. For each mode, we then determine the distance between the basin edges in the positive and negative directions, $\delta_{\mu}$.

We define a quantity which we call the ``compact volume'', $V_{N}^{\rm{comp}}(n)$, of this basin $n$, as the product of the distances $V_{N}^{\rm{comp}}(n) \equiv \prod_{\mu= d+1}^{Nd} \delta_{\mu}$ where we have omitted the $d$ trivial $\omega=0$ modes. We emphasize that this quantity is an estimate of the basin volume \textit{only} if the basin is uncontorted. In defining $V_N^{\rm{comp}}(n)$, we approximate the basin as a hyper-rectangle in $Nd$-dimensional space. The axes of this hyper-rectangle are chosen to be the eigenvectors of the Hessian which form an orthonormal basis in coordinate space and correspond to the different curvatures around the minimum of the equilibrium configuration potential energy surface. For each basin, we then define the contortion as
\begin{equation}
\Xi_{N}(n) \equiv V_{N}(n)/V_{N}^{\rm{comp}}(n).
\label{contortiondef}
\end{equation}
If $\Xi_N(n)$ is very large, then the compact volume is only a small fraction of the true volume. We observe that $\Xi_N(n)$ varies enormously, even between basins with nearly equal volumes.

\begin{figure}
    \centering
    \includegraphics[width=1\linewidth]{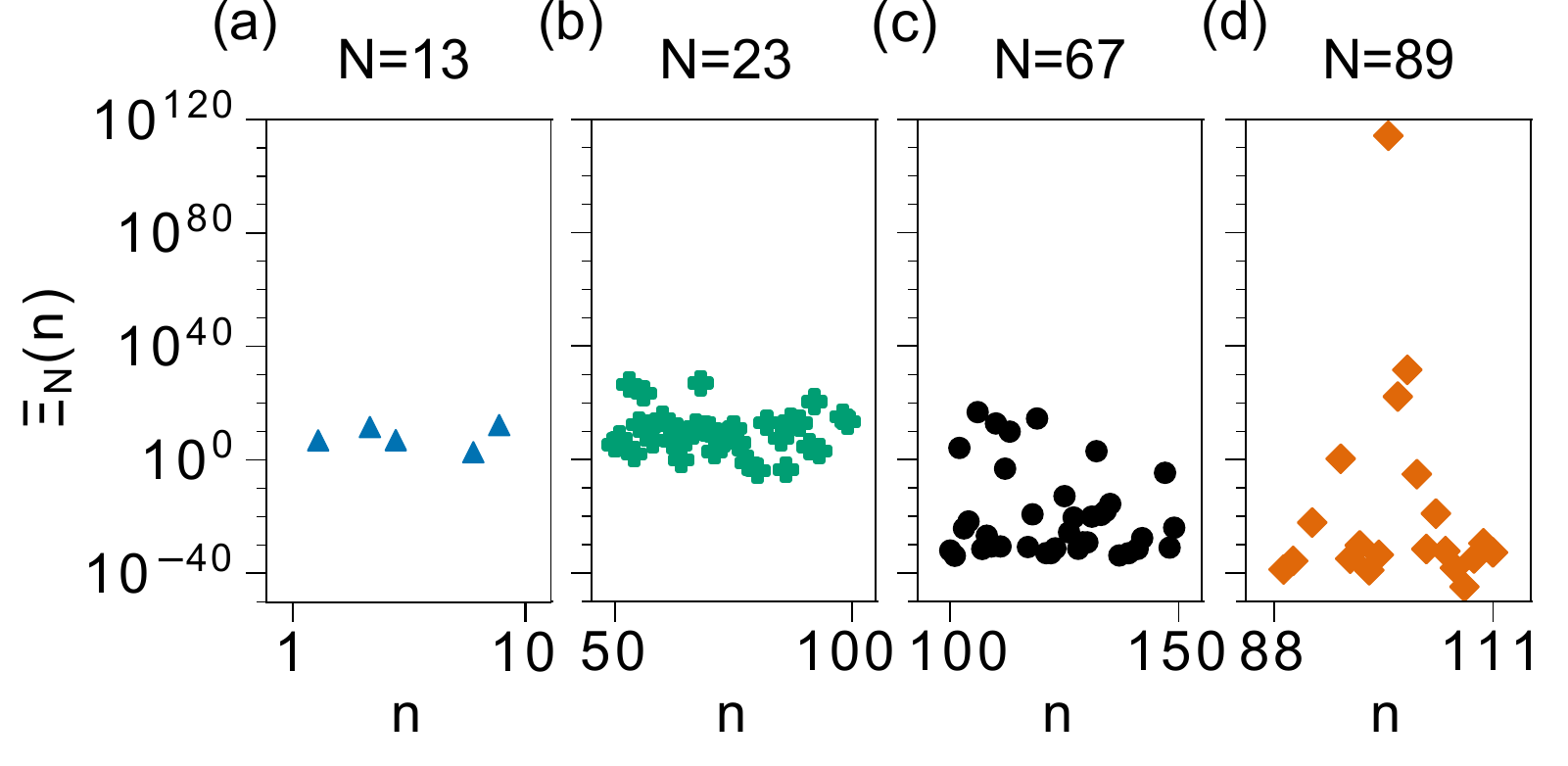}
    \caption{Contortion, $\Xi_N(n) \equiv V_N(n)/V_N^{\rm{comp}}(n)$, versus $n$ for systems with four sizes in $d=3$. For (a) $N=13$ in range $1 \leq n \leq 10$, (b) $N=23$ in range $50 \leq n \leq 100$, (c) $N=67$ in range $100 \leq n \leq 150$, and (d) $N=89$ in range $88 \leq n \leq 111$. There are enormous fluctuations in $\Xi$, in some cases spanning more than $50$ decades, even between adjacent values of $n$. As $N$ increases, the spread in $\Xi$ also increases. The compact volume, $V_N^{\rm{comp}}(n)$, is computed from the hyper-rectangle approximation described in the text. 
    }
    \label{fig:compactness-3D}
\end{figure}

Fig.~\ref{fig:compactness-3D} shows $\Xi_N(n) \equiv V_N(n)/V_N^{\rm{comp}}(n)$ for the $N=13$, $23$, $67$, and $89$ data over varying ranges of $n$. If the basins were uncontorted, one might expect this ratio to remain approximately constant ($\approx 1$) as a function of $n$. However, as shown in the figure, the values of this ratio for basins with similar values of $n$ can vary by many orders of magnitude. These fluctuations grow with system size, $N$; for $N=89$, they span more than $160$ orders of magnitude. This variation is far too large to be consistent with a simple hyper-rectangle approximation to the basin volumes.

We can regard $\Xi$ as a measure of how contorted the basins are. It is striking that a quantity that is relatively easy to compute can vary by more than $160$ orders of magnitude between basins in the landscape. One possible picture that emerges is that the surface of each basin is highly ramified, with overhangs and re-entrant regions; in this picture, some basins are much smoother than others. Another possibility is that the basin surface remains smooth but is not aligned with the axes of the hyper-rectangle determined near the minimum of the potential energy surface. 

One way to distinguish between these possibilities is to examine long, straight trajectories that start at the equilibrium configuration of each basin and count how many times the system falls back into its original basin. We have implemented this procedure for analyzing paths emerging from a few sample basins and find that overhanging regions do indeed exist. Following a straight-line path starting at a basin minimum brings us repeatedly out of and then back again into the initial basin. Establishing whether there is a correlation between $\Xi$ and the number of such re-entrant paths, as suggested above, would require a systematic study of many paths across many basins, which is beyond the scope of the present paper.
While the complexity of high-dimensional surfaces has certainly been recognized previously, $\Xi$ is a simple and useful measure of how complex the basin boundary can be. The fact that it varies so much between basins with nearly identical volumes is remarkable and difficult to ignore.

\section{Discussion: Relevance to memory formation}

Our results are also relevant to the nature of memories produced by cyclic shear in disordered packings. Experiments~\cite{Keim14,mukherji2019,lindeman2025} and simulations~\cite{fiocco2014encoding,Royer15,lavrentovich2017period,keim2021multiperiodic,arceri2021marginal} have shown that, after only a few cycles of cyclic shear, a jammed packing can become periodic so that it repeatedly visits the same states in each cycle. This surprising behavior has been attributed to the presence of hysterons or defects in the material that undo themselves when the shear direction is reversed~\cite{mungan2019networks,mungan2019cyclic,lindeman2021multiple,van2021profusion,lindeman2025minimal}. The present work suggests a complementary approach to understanding this behavior, one that focuses not on the microscopic instabilities themselves but on the statistics of catchment basins in the energy landscape.

If the total number of distinct basins, $n_{\rm tot}$, is vast, it is difficult to imagine how a system can return periodically to the same basin after only a few cycles of training.  However, we have found a very broad distribution of basin volumes some of which are much larger than others. Once the system has reached one of these very large basins, it is much more likely to return to it in subsequent cycles. Once two trajectories reach the same basin, they will subsequently follow the same deterministic path in the quasistatic regime. This provides a mechanism by which states can be funneled into larger and larger basins, as suggested by Gao \textit{et al.}~\cite{gao2009geometrical}.

\begin{figure}
    \centering
    \includegraphics[width=1\linewidth]{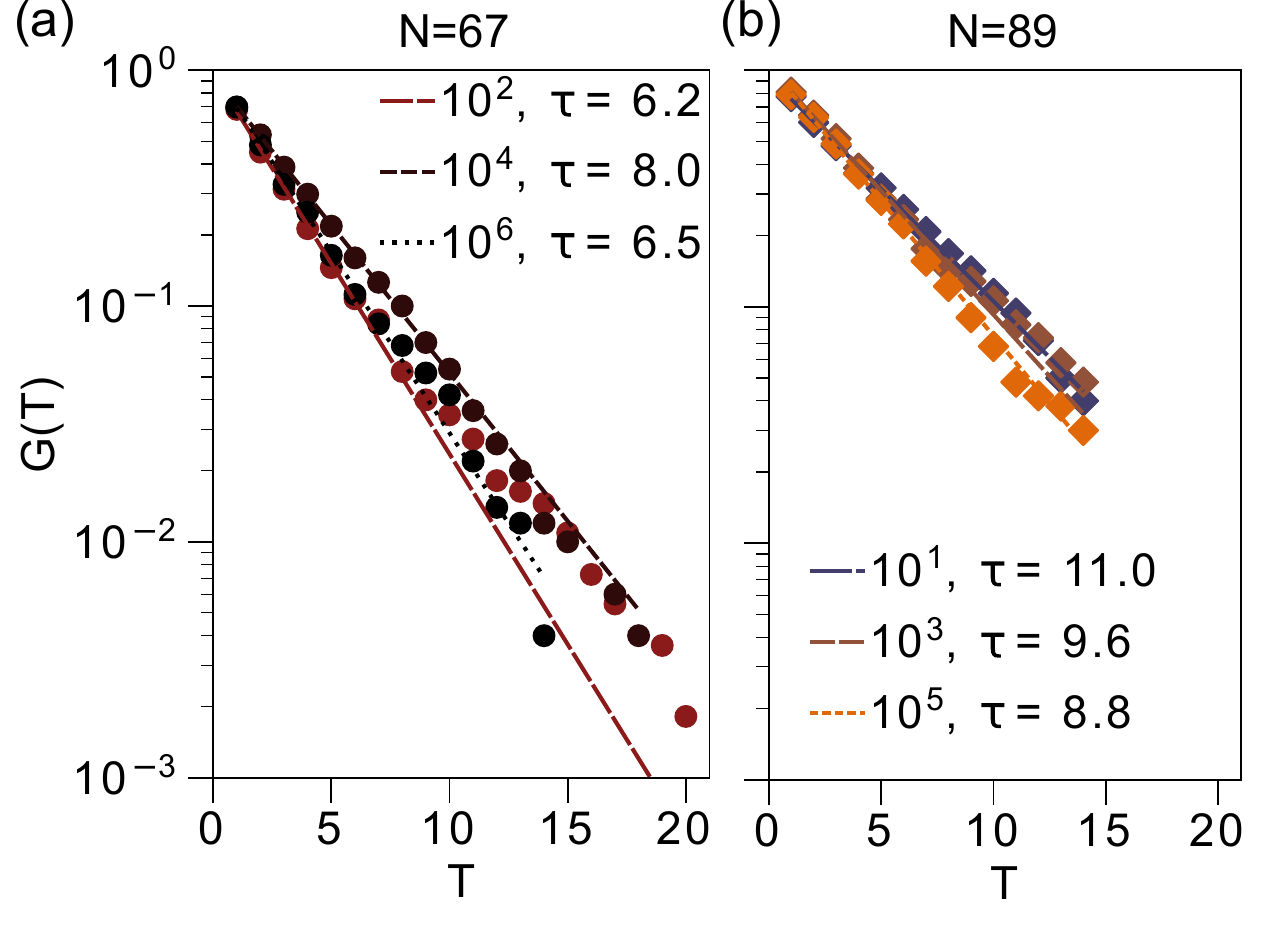}
    \caption{The complementary cumulative probability of reaching the first decade in $T$ steps via a random walk between neighboring basins in $d=3$ for $N=67$ and $N=89$. Each symbol represents data from a different starting decade. The straight-lines are fits to an exponential decay, $G(T) = 10^{-T/\tau}$, with $\tau$ fit separately for the data from each starting decade. The fitted values of $\tau$ are shown in the legends. The average decay for all three curves is $\langle\tau \rangle = 6.9$ for $N=67$ and $\langle \tau \rangle = 9.7$ for $N=89$. 
    }
    \label{fig:timetolargestbasins}
\end{figure}

To make this argument quantitative, we start in an arbitrarily chosen basin $n$, displace the configuration a distance $\delta$ in a random direction, and minimize the energy from the resulting configuration to identify the neighboring basin $n_{\rm nn}$ that we land in. We then repeat the procedure from $n_{\rm nn}$, moving in a new random direction and minimizing again to reach $n_{\rm nnn}$, and so on ($n \rightarrow n_{\rm nn} \rightarrow n_{\rm nnn} \rightarrow \cdots$), until we arrive at a basin in the top decade of $n$. Averaging over many initial basins gives the first-passage distribution $g(t)$, the probability that the system first reaches a basin in the top decade after exactly $t$ steps. In Fig.~\ref{fig:timetolargestbasins}, we plot the complementary cumulative distribution function, $G(T) \equiv (1-\sum_1^T g(t))$ versus T. Each data set corresponds to starting from basins in a different decade of $n$. On this semi-log plot, a straight line implies the exponential form $G(T) = 10^{-T/\tau}$, where $\tau$ is the number of steps per decade of decay and depends on the system size $N$. The slopes do not vary appreciably between starting decades, so $\tau$ is essentially independent of how small the initial basin is.

To understand this data, we assume that the probability of finding a basin of size $y=\log n$ is constant. This was demonstrated to be a good approximation in Fig.~\ref{fig:neighboring-basins}.  In each step, every decade is therefore equally likely to be chosen with probability $1/N_{\mathrm{dec}}$ (where $N_{\mathrm{dec}} \equiv \log_{10} n_{\rm{tot}}$ is the total number of decades). Thus the probability to reach a basin in the top decade (\textit{i.e.}, one of the ten largest basins) for the first time after $t$ steps is $g(t) = \frac{1}{N_{\mathrm{dec}}}(1-\frac{1}{N_{\mathrm{dec}}})^{t-1}$. Therefore, 
\begin{equation}
G(T) = 1-\sum_{t=1}^T \frac{1}{N_{\mathrm{dec}}}(1-\frac{1}{N_{\mathrm{dec}}})^{t-1}
=(1-\frac{1}{N_{\mathrm{dec}}})^T 
\label{G(T)}
\end{equation}
is an exponential function of form $G(T) = 10^{-T/\tau}$, with $N_\mathrm{dec}=(1-10^{-1/\tau})^{-1}$. This form is represented by the straight lines in Fig.~\ref{fig:timetolargestbasins} and is consistent with the data. 

For the system with $N=67$ in $d=3$, we have identified a few of the basins that form periodic orbits after cyclic shearing to $\gamma = 0.5$ in increments of $\delta \gamma = 2 \times 10^{-5}$.  (With such a large shear, many basins must be visited in both the forward and backward directions.) For example, when we started out in basin $n=10,000$, we found that after one shear cycle, the system ends up in a periodic orbit that starts and ends in basin $n=2$. More interesting, perhaps, is that when starting from basin $n=100$, it takes \textit{two} cycles to reach the periodic orbit that starts and ends in basin $n=3$. We also find orbits that take multiple periods before they repeat themselves. These last two examples would not have been possible with non-interacting hysteron models which must reach the periodic orbit in only one cycle and which must have orbits with periods of only one cycle~\cite{lindeman2021multiple,lindeman2025}.

In contrast to explanations based primarily on local rearrangements, this argument emphasizes the statistics and connectivity of catchment basins in the energy landscape. Several caveats to this argument should be emphasized. We have considered only monodisperse, identical particles, so our analysis is insensitive to particle permutations. If particles exchange positions during a shear perturbation, the resulting configuration would not be considered to belong to the same basin if the particles were distinguishable. In addition, we have sampled neighboring configurations using random perturbations, whereas in the case of cyclic shear the boundary motion is deterministic. This may therefore introduce additional correlations between the neighboring basins visited. Despite these caveats, the broad distribution of catchment-basin volumes should still enhance the mechanism of funneling toward larger basins~\cite{gao2009geometrical}.

\section{Conclusions}

We have shown that, for monodisperse packings of soft particles, ranking the basins of the energy landscape by their volumes reveals Zipf-like distributions in both $d=3$ and $d=2$. This is notable because two-dimensional packings are essentially polycrystalline systems perturbed by defects, whereas three-dimensional packings exhibit more evident structural disorder. This similarity suggests that the distributions we observe may be a consequence of the high dimensionality of the energy landscape rather than of the specific geometry of the packings. From this perspective, the energy landscape of jamming may be one example of a broader class of high dimensional landscapes with similar basin volume statistics.  This would then present a possible way that nature allows systems to organize preferentially into certain classes of structures.

We emphasize that the basin-volume distributions can also be fit essentially equally well by several functional forms, including a log-normal distribution. However, the part most relevant for the physical properties of the system is the small-$n$ regime, corresponding to the largest basin volumes $V_n$. Even for our largest systems, $N=151$ in $d=3$ and $N=509$ in $d=2$, a sizable fraction of all random initial conditions fall within the approximately power-law regime sampled here. Moreover, for the systems in which we have sufficient data, the overwhelming contribution to the total sampled volume comes from this high-volume portion of the distribution. For instance, for $N=67$ in $d=3$, our sampled basins cover more than $98\%$ of the total phase space volume of the system. Similarly, for $N=197$ in $d=2$, they cover more than $97\%$ of the volume. Thus, whether the full distribution eventually develops a log-normal peak outside our sampled range, at smaller basin volumes and larger $n$, does not alter the central conclusion that the quasistatic behavior of these systems is dominated by the largest basins.

Access to many decades of basin volumes provides an opportunity to investigate quantitatively properties of the energy landscape that would otherwise be difficult to explore. In particular, we find no correlation between the volume of a basin and the volumes of its neighbors. Likewise, we are able to characterize the shapes of basins in the $Nd$-dimensional landscape and show that they are highly contorted in the sense defined earlier. 

Our focus on monodisperse indistinguishable particles allowed us to sample the volume distributions over a wide range in remarkably large systems. When polydispersity is introduced, the permutation of the particles in one given state may also change the set of interacting (\textit{i.e.}, overlapping) neighbors. That is, there are more complicated interactions occurring that do more than merely permute the particle labels. There will be small perturbations in the geometrical structure of the configurations depending on which permutation is chosen. Such rearrangements will alter the basins and could influence how the catchment basin statistics affect the behavior of disordered solids.  The monodisperse case studied in the present paper serves as the symmetry point about which such perturbations can be studied cleanly.

\section{Methods}
For all system sizes and dimensions, we minimized the energy of each configuration to quad precision using a GPU-based implementation of the FIRE algorithm~\cite{morse2014geometric}. To maximize the sampling number for  $N=67$, $d=3$, we supplemented this with a CPU-based Fortran implementation of FIRE with double precision. 
\subsection{Fortran algorithm parameters for FIRE}
At each step in the Fortran code, the particle velocities $\mathbf{v}$ were %modified according to the FIRE prescription by 
projected onto the force directions. The adaptive time step was increased by a factor of 1.1 after five consecutive steps with positive power $P=\mathbf{F}\cdot\mathbf{v}$, while the mixing parameter $\alpha$ was reduced by a factor of 0.99. When $P\le0$, the velocities were reset to zero, the time step was reduced by a factor of 0.5, and  $\alpha$ was restored to its initial value of 0.1. The maximum allowed time step was set to ten times the initial time step. Minimization was continued until the dimensionless force residual, defined as $\max |F_i| R$, fell below $10^{-13}$, where $R$ is the particle radius and $|F_i|$ is the magnitude of force acting on particle $i$.  A maximum of $10^6$ FIRE iterations was allowed for each minimization. %The code is available in the zenodo data recovery. 

\subsection{FIRE versus steepest descent} It is well known that different minimization algorithms can lead to different ground states~\cite{wales1982basins,asenjo2013visualizing,nishikawa2022relaxstion,bautista2026}.  To check whether minimization using FIRE introduces a bias to our measurements of the basin-volume distributions, we compare FIRE with steepest descent when starting from identical initial random configurations. As shown in Fig.~\ref{fig:comparison-algo} in the Supplemental Information, for both $N=13$ and $N=67$ in $d=3$, we find no significant discrepancies in between the two minimization procedures. While the two algorithms can lead to different basins, the distributions are essentially within counting statistics of each other. We conclude that the initial `kick' inherent in the FIRE algorithm is itself sufficiently random so that it does not grossly change the probability of finding basins of different sizes. Thus, over the range that we can access with such comparisons, we find similar results. 

\section*{Acknowledgments}
We are particularly grateful to Justin Burton for the use of his Fortran FIRE energy minimization code.  We thank Eric Corwin, Daan Frenkel, Nathan Keim, Dov Levine, Andrea Liu, Corey O'Hern, Joseph Paulsen, Vincenzo Vitelli, Thomas Witten, and Ning Xu for stimulating discussions over many years.   
This work was supported at the University of Chicago by the NSF Center for Living Systems grant PHY-2317138. 
%US Department of Energy, Office of Science, Basic Energy Sciences, under Grant DE-SC0020972. Computational resources were provided by the NSF MRSEC program NSF-DMR 2011854 
and the University of Chicago Research Computing Center.
This work used CPUs and GPUs at NCSA Delta (UIUC) through allocations PHY240083 and PHY240137 from the Advanced Cyberinfrastructure Coordination Ecosystem: Services \& Support (ACCESS) program, which is supported by U.S. National Science Foundation grants \#2138259, \#2138286, \#2138307, \#2137603, and \#2138296.\cite{ACCESS}

\section*{Author Contribution}
V.F.H. and Z.L. contributed equally to this work. S.R.N., V.F.H., and Z.L. conceived the study and developed the analysis. Z.L. led the numerical simulations, data analysis, and visualizations in the text. V.F.H. contributed to the initial analysis of basin-volume statistics in three dimensions and to the interpretation of the results. S.R.N. supervised the project and led the writing of the manuscript. All authors discussed the results, reviewed the manuscript, and contributed to its final form.

\section*{DATA AVAILABILITY}
The data that support the findings of this article are included in the main text and Supplemental Material and are available upon request. The access to the GPU-based code will also be available upon request. The parameters used for the CPU-based FORTRAN code for the FIRE algorithm are included in the Supplemental Material.

\bibliography{main} %Produces the bibliography via BibTeX.
\newpage
\clearpage

\renewcommand{\thefigure}{S\arabic{figure}}
\renewcommand{\thetable}{S\arabic{table}}
\renewcommand{\theequation}{S\arabic{equation}}
\renewcommand{\thepage}{S\arabic{page}}
\setcounter{figure}{0}
\setcounter{table}{0}
\setcounter{equation}{0}
\setcounter{page}{1}

\section*{Supplementary Information}

%In the CPU-based code, particle positions were iteratively updated according to the forces derived from the energy. 
\subsection*{FIRE versus steepest descent minimization algorithms}

We use the FIRE algorithm to minimize the energy and obtain the stable configuration corresponding to each set of random initial particle positions. We use this fast minimization algorithm in order to perform the large number of trials, $m_{\rm{tot}}$, needed to sample phase space accurately.

\begin{figure}
    \centering
    \renewcommand{\thefigure}{S\arabic{figure}}
    \includegraphics[width=1\linewidth]{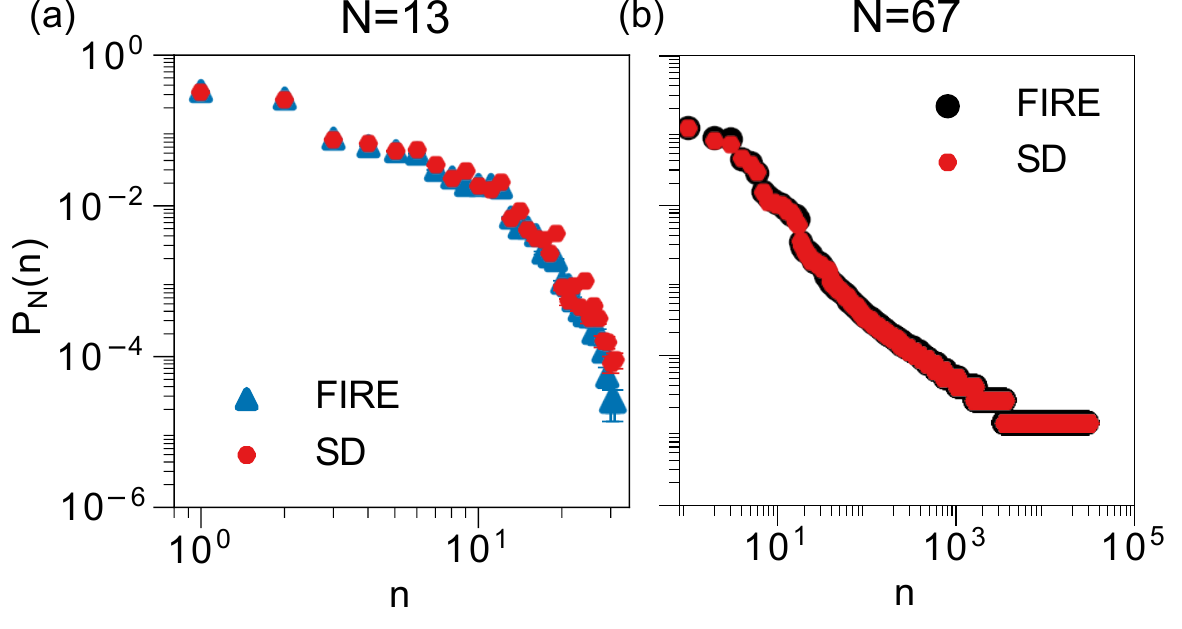}
    \caption{Comparison of catchment basin volumes determined by FIRE and steepest descent algorithms. (a) $P_{13}(n)$ with sample size $m_{\rm{tot}} = 2 \times 10^5$. The distribution from steepest descent is sorted based on the rank from FIRE algorithm. (b) $P_{67}(n)$ with sample size $m_{\rm{tot}} = 79,829$. Both algorithms give similar results. Symbols for FIRE and steepest descent (SD) are defined in the legend. The FIRE symbols are chosen to be slightly larger for visibility.}
    \label{fig:comparison-algo}
\end{figure}

To test whether the FIRE algorithm introduces bias into our $P_N(n)$ distributions, we also determined the basin-volume distributions using a steepest descent algorithm (SD). The results are shown in Fig.~\ref{fig:comparison-algo}. Over the range in which the comparison can be made, we observe only very small deviations between the two sets of results. This indicates that the FIRE algorithm does not significantly bias the statistics of our distributions.

\subsection*{Phase space volumes sampled by simulations} 

\begin{figure}
    \centering
    \includegraphics[width=1\linewidth]{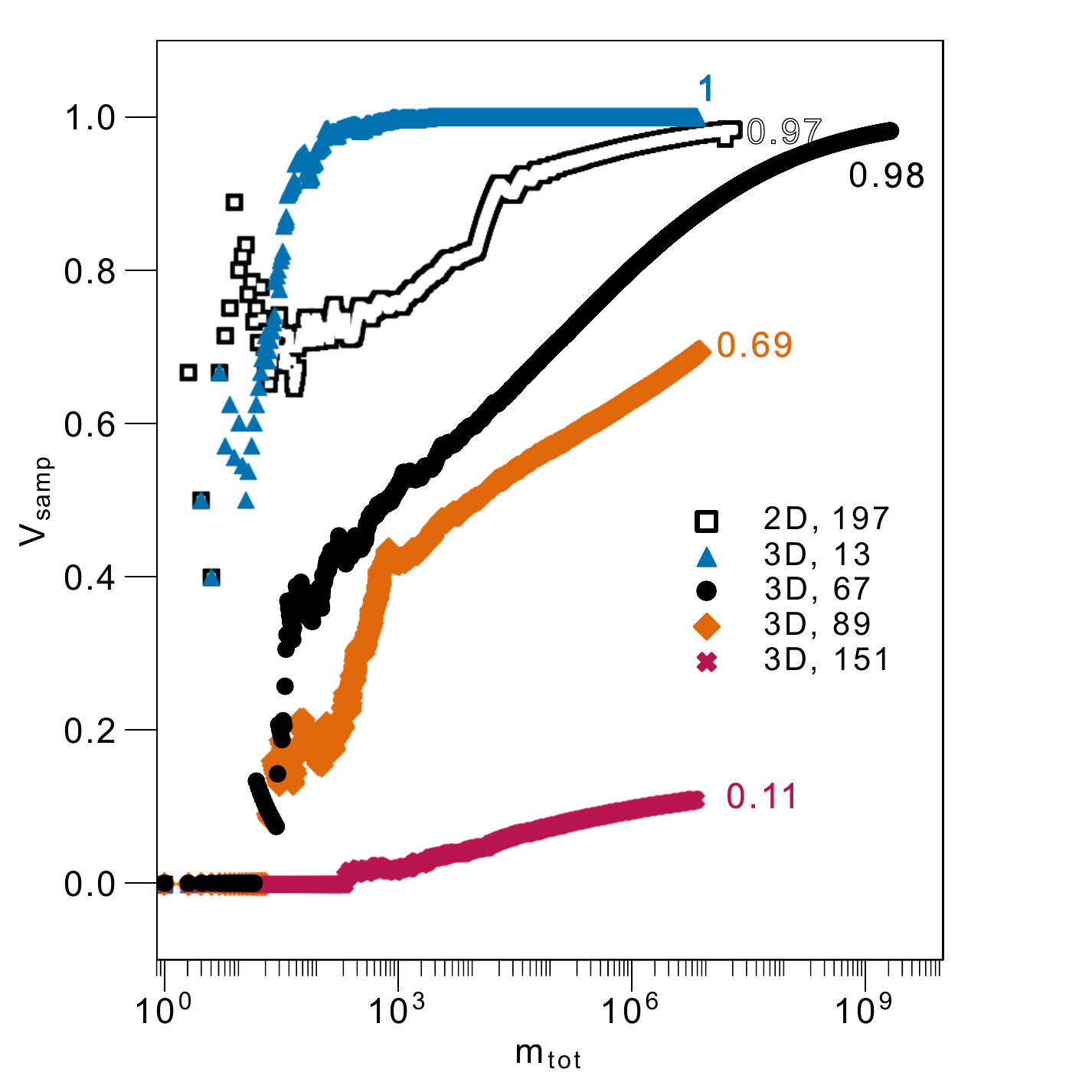}
    \caption{Total sampled phase space volume, $V_{\rm{samp}}$, versus the total number of trials, $m_{\rm{tot}}$, for $N=13$, $67$, $89$, and $151$ in $d=3$, and for $N=197$ in $d=2$; symbols are defined in the legend. For these distributions, the final sampled volumes are $V_{\rm{samp}} > 0.99$, $0.98$, $0.69$, and $0.11$ for $N=13$, $67$, $89$, and $151$ in $d=3$, respectively, and $V_{\rm{samp}} > 0.97$ for $N=197$ in $d=2$.    
    }
    \label{fig:phase-space-volume}
\end{figure}

Fig.~\ref{fig:phase-space-volume} shows the estimated fraction of the total phase space volume sampled in our simulations, restricted to basins with $m_n \geq 2$, for systems of different sizes, $N$, in both $d=3$ and $d=2$. As $N$ increases, it becomes progressively more difficult to sample a large fraction of the phase space. For $N=67$ in $d=3$ and $N=197$ in $d=2$, we sampled fractions $V_{\rm{samp}} > 0.98$ and $V_{\rm{samp}} > 0.97$, respectively. Even for $N=151$ in $d=3$, we sampled $V_{\rm{samp}} > 0.11$. Thus, even in this large system, more than $10\%$ of initial conditions fall into basins that were sampled at least twice in our trials.

\subsection*{Power-Law versus Log-Normal Fit}

Previous studies suggested that the density of catchment-basin volumes obeys a log-normal distribution~\cite{xu2011direct,asenjo2014numerical,paillusson2015devising,martiniani2017numerical,casiulis2023when,frenkel2013other} of the form,
\begin{equation}
    P_N(V) = \frac{1}{V\sigma \sqrt{2\pi}} \exp(-\frac{(\ln V - \mu)^2}{2\sigma^2}).
\label{eq:Lognormal}
\end{equation}
where $\mu$ and $\sigma$ are respectively the mean and standard deviation of $\ln P_N(n)$. In order to find the distribution $P_N(n)$ that corresponds to the log-normal distribution, we use Eq.~\ref{Pisvolume},
$V(n) = P_N(n)$ to derive
\begin{equation}
n = -\frac{\exp(-\mu-\sigma^2/2)}{2} \erf (\frac{\ln V(n)- \mu}{\sqrt{2}\sigma})+const.
\label{Lognormalderiv}
\end{equation}

%Detailed deviation attached below:
%\begin{equation*}
 %   \begin{split}
 %   &<V>=P_N(V)VdV\\  
 %   &V_{total}=n_{max}<V>=n_{max}P_N(V)VdV=P_N(n)dn=Vdn\\
 %   &\Rightarrow n_{max}\int_V^{V_{max}} P_N(V')dV'=\int_{n}^1dn=1-n\\
 %   &\Rightarrow n=-n_{max}\erf(\frac{\ln V-\mu}{\sqrt{2}\sigma})+const
 %   \end{split}
%\end{equation*}}
%The mean of lognormal distribution $<V>=\exp(\mu+\sigma^2/2)$, so $n_{max}=\exp(-\mu-\sigma^2/2)$.

%\begin{equation*}
 %   \begin{split}
 %   &\Rightarrow n=-n_{max}\erf(\frac{\ln V-\mu}{\sqrt{2}\sigma})+const\\
 %   &= -\exp(-\mu-\sigma^2/2)\erf(\frac{\ln V-\mu}{\sqrt{2}\sigma})+const
 %   \end{split}
%\end{equation*}

We have been careful not to argue that any one functional form provides a better fit to $P_N(n)$ than another. No form is perfect. Over the range accessible to our sampling, which in some cases extends over eight decades in $n$, it is not clear whether a log-normal fit,
is significantly better or worse than a simple power-law form, $P_N(n) = A_N n^{-\alpha}$ with $\alpha \approx 1$ and $A_N$ determined by normalization of the distribution function, as found in typical Zipf-like distributions. 
\begin{figure}[h]
    \centering
    \includegraphics[width=1.1\linewidth]{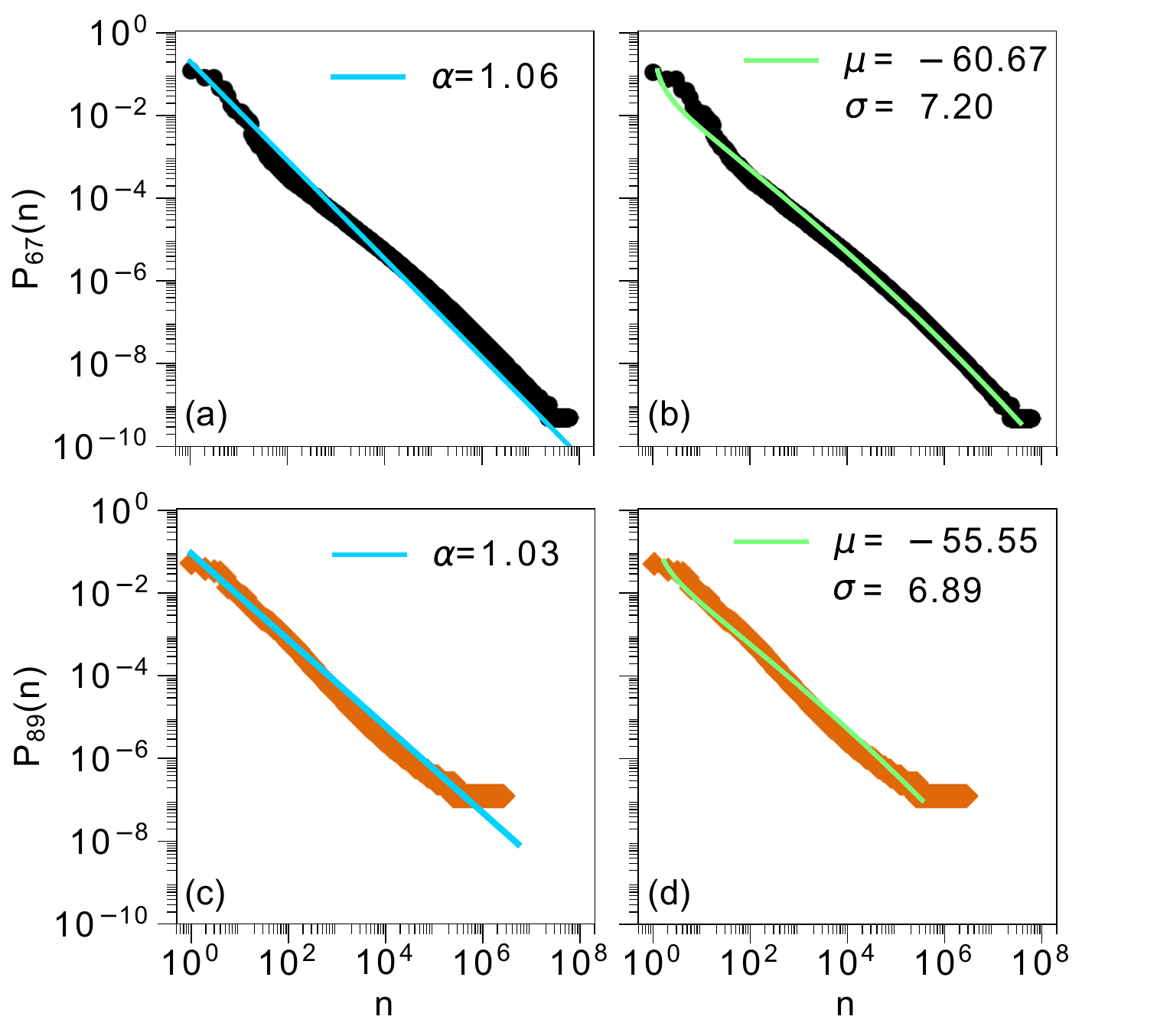}
    \caption{Comparison between power-law and log-normal fits for three-dimensional jammed packings with different sizes. The power-law exponents for the fits in (a) for $N=67$ and (c) for $N=89$ are shown in the legends. For these fits, the prefactor is determined by requiring the total probability to sum to unity. In (b) and (d), the legends show the log-normal fit parameters $\mu$ and $\sigma$ in Eq.~\ref{eq:Lognormal}}.
    \label{fig:comparison-3D}
\end{figure}

\begin{figure}
    \centering
    \includegraphics[width=1\linewidth]{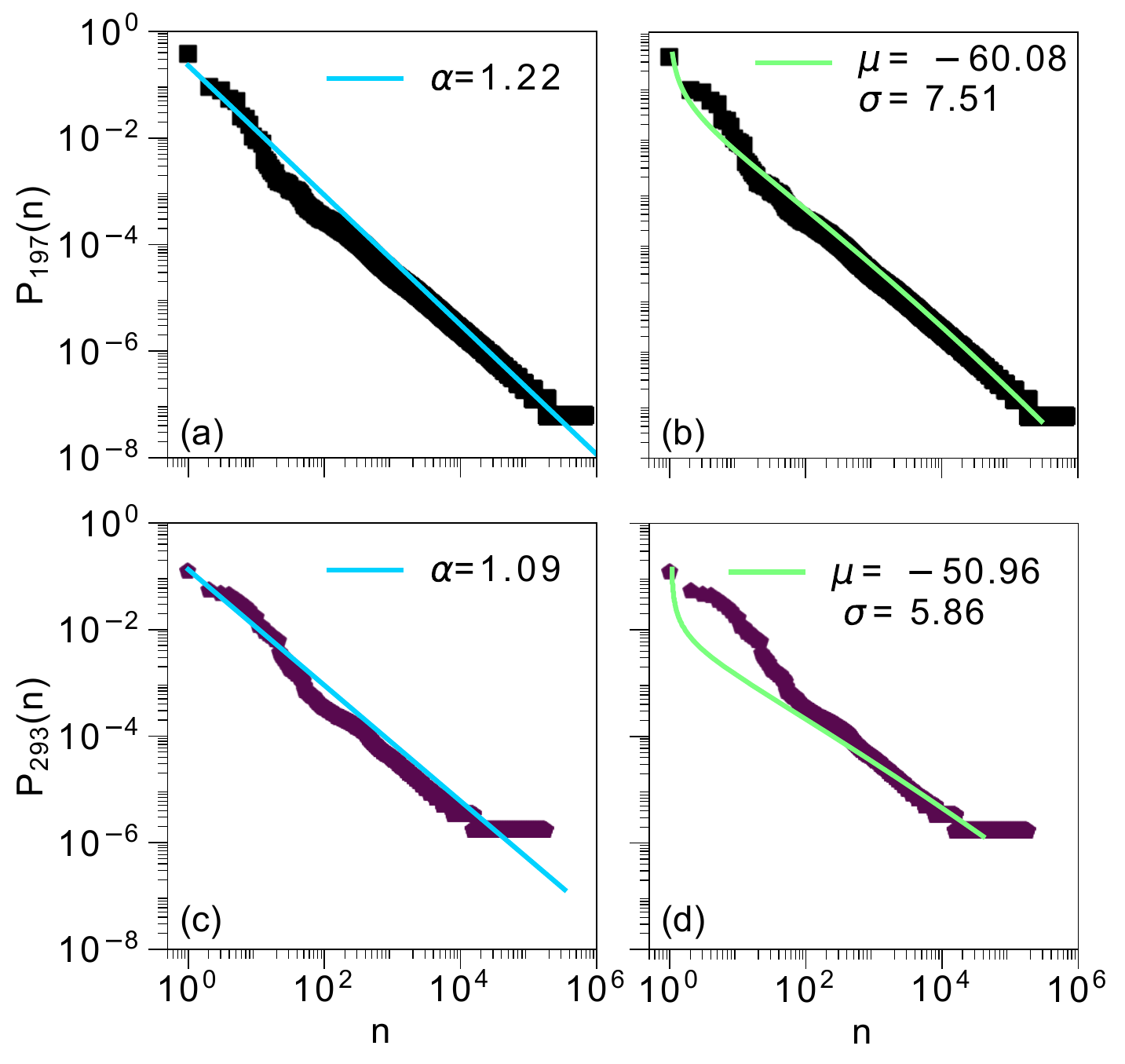}
    \caption{Comparison between power-law and log-normal fits for two-dimensional jammed packings with different sizes The power-law exponents for the fits in (a) for $N=197$ and (c) $N=293$ are shown in the legends. For these fits, the prefactor is determined by requiring the total probability to sum to unity. In (b) and (d), the legends show the log-normal fit parameters $\mu$ and $\sigma$ in Eq.~\ref{eq:Lognormal}.}
    
    \label{fig:comparison-2D}
\end{figure}

We compare these two fitting forms for $d=3$ and $d=2$ in Figs.~\ref{fig:comparison-3D} and \ref{fig:comparison-2D}, respectively. The feature that would most clearly distinguish the two forms is the presence of a peak. However, there is no evidence of such a peak in the region that we have been able to sample. If such a peak is physically important, it is surprising that it does not appear in $d=3$ for our $N=67$ data, for which we have sampled more than $0.98$ of the total phase space volume or in $d=2$ for our $N=197$ data where we have sampled $0.97$ of the total volume.

\subsection*{Potential energy of stable configurations versus basin volume}

\begin{figure}
    \centering
    \includegraphics[width=1\linewidth]{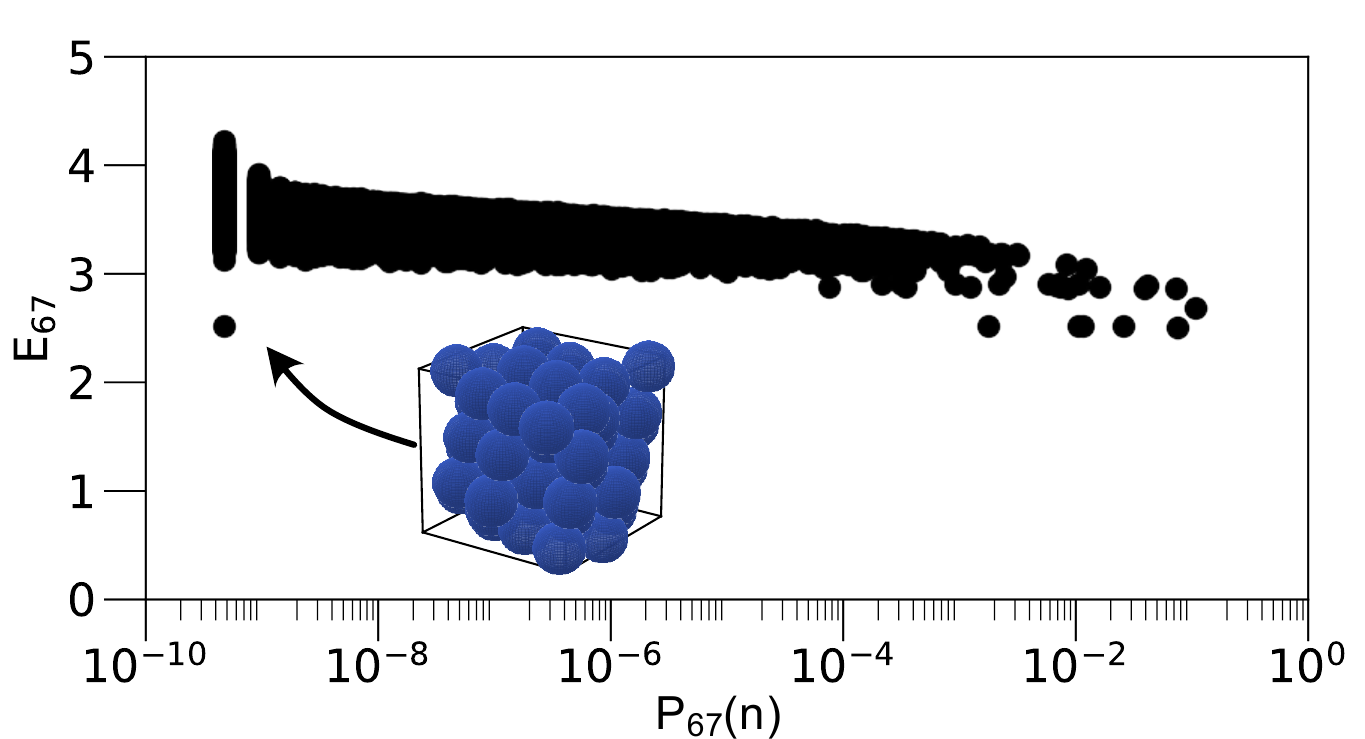}
    \caption{Potential energy of stable configurations, $E_{N}(n)$, versus catchment basin volume, $P_{N}(n)=V_{N}(n)$ for packings with $N=67$ particles. The largest basin does not correspond to the lowest potential energy configuration. The basin with the third lowest potential energy was found only once in our sampling. The inset shows this configuration, which appears to be partially crystallized.}
    \label{fig:energy-p-67}
\end{figure}

One might have expected the volume of a basin, $V_N(n)=P_N(n)$, to be correlated with the total potential energy of the stable configuration at its minimum, 
\begin{equation}
E(n) = \frac{1}{2} \sum_{\rm{i \neq j}} E_{i,j}
\label{Energ}
\end{equation}
where $E_{i,j}$ is given by Eq.~\ref{potential} and the sum is over all pairs of particles in configuration $n$.  Fig.~\ref{fig:energy-p-67} shows $E_{N}(n)$ versus $P_{N}(n)=V_{N}(n)$ for $N=67$. Except for the very largest catchment basins, the spread of the energy $E$ in each increment of volume is larger than any small systematic variation of the average energy with volume.  

In particular, the largest volume basin does not have the lowest potential energy. Interestingly, the configuration with the third lowest observed potential energy was sampled only once in the $m_{\rm{tot}} = 2 \times 10^9$ trials. It therefore has a very small catchment volume. It corresponds to a partially crystallized configuration, as shown in the inset.

\subsection*{Fluctuations in $P_N(n)$ in $d=2$}

\begin{figure}
    \centering
    \includegraphics[width=1\linewidth]{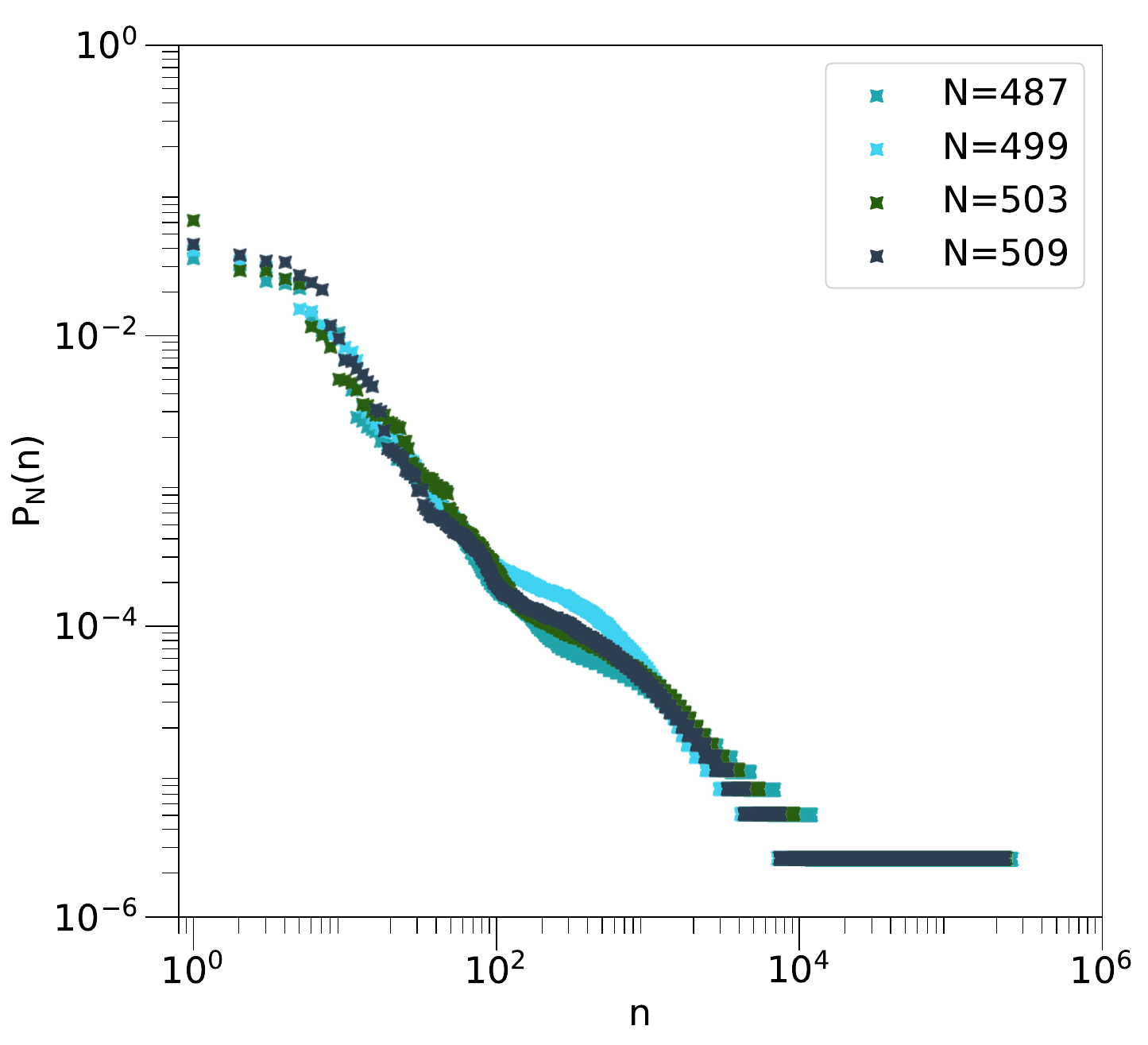}
    \caption{Distribution of catchment-basin sizes for four system sizes showing that features do not move systematically with $N$. In $d=2$, there are significant fluctuations in $P_N(n)$ even when $N$ varies within only a small range. The fluctuations do not appear to correlate with $N$. 
    }
    \label{fig:fluctuations-2D}
\end{figure}

As $N$ increases in $d=2$, the distributions shown for $N=197$ and $N=293$ in Fig.~\ref{fig:vol-dist-2D} of the main text begin to develop structure and are no longer completely straight on a log-log plot. However, this structure does not vary systematically with increasing $N$ over the range that we can probe. Fig.~\ref{fig:fluctuations-2D} shows that varying $N$ by small increments between $N=487$ and $N=509$ appears to shift the structure in non-systematic ways.

%\subsection*{Steps to reach largest basins}

%We wish to see how long it will take random hopping between nearest-neighbor basins to reach one of the few particularly large basins. This is particularly important for determining how the distribution of basin volumes affects the memory of a cyclically sheared jammed packing. We determined that Following the procedure we used above in Section IIE 

%We can compute how many steps $n \rightarrow n_{nn} \rightarrow n_{nnn}$ \textit{etc.}  does it take, moving in a random direction from an initial randomly chosen basin to its neighbor ($n_{nn}$), and then from that neighbor to one of its neighbors ($n_{nnn}$ \textit{etc.}) until the system reaches a basin with one of the top ten largest values.  That is we wish to see if random hopping between basins will lead to eventual placement in a particularly large basin.  

%This shows consistency of the distributions $P_N(n)$ remaining as a power law out to far beyond where we have been able to determine it by straight-forward measurement. We can compare the average value of $\tau$ for each N with the value expected from the argument that each decade in $n$ is equally probable. For $N=67$, the measured value of $\tau$ predicts that there are a total of $3.6$ decades and for $N=89$ it predicts $4.8$ decades.

\include{SI}

\end{document}